\documentclass[twocolumn,floatfix,eqsecnum,showpacs,superscriptaddress]{revtex4}
\usepackage{graphicx}
\usepackage{amsmath}
\usepackage{bm}
\begin{document}

\title{Extremal transmission at the Dirac point of a photonic band structure}
\author{R. A. Sepkhanov}
\affiliation{Instituut-Lorentz, Universiteit Leiden, P.O. Box 9506, 2300 RA Leiden, The Netherlands}
\author{Ya.\ B. Bazaliy}
\affiliation{Instituut-Lorentz, Universiteit Leiden, P.O. Box 9506, 2300 RA Leiden, The Netherlands}
\affiliation{Department of Physics and Astronomy, University of South Carolina,
Columbia, SC 29208, USA}
\author{C. W. J. Beenakker}
\affiliation{Instituut-Lorentz, Universiteit Leiden, P.O. Box 9506, 2300 RA Leiden, The Netherlands}
\date{March 2006}
\begin{abstract}
We calculate the effect of a Dirac point (a conical singularity in the band structure) on the transmission of monochromatic radiation through a photonic crystal. The transmission as a function of frequency has an extremum near the Dirac point, depending on the transparencies of the interfaces with free space. The extremal transmission $T_{0}=\Gamma_{0} W/L$ is inversely proportional to the longitudinal dimension $L$ of the crystal (for $L$ larger than the lattice constant and smaller than the transverse dimension $W$). The interface transparencies affect the proportionality constant $\Gamma_{0}$, and they determine whether the extremum is a minimum or a maximum, but they do not affect the ``pseudo-diffusive'' $1/L$ dependence of $T_{0}$.
\end{abstract}
\maketitle

\section{Introduction}
\label{intro}

In a two-dimensional photonic crystal with inversion symmetry the band gap may become vanishingly small at corners of the Brillouin zone, where two bands touch as a pair of cones. Such a conical singularity is also referred to as a Dirac point, because the two-dimensional Dirac equation has the same conical dispersion relation. In a seminal work \cite{Rag06}, Raghu and Haldane investigated the effects of broken inversion symmetry and broken time reversal symmetry on the Dirac point of an infinite photonic crystal. Here we consider the transmission of radiation through an ideal but finite crystal, embedded in free space.

As we will show, the proximity to the Dirac point is associated with an unusual scaling of the transmitted photon current $I$ with the length $L$ of the photonic crystal. We assume that $L$ is large compared to the lattice constant $a$ but small compared to the transverse dimension $W$ of the crystal. For a true band gap, $I$ would be suppressed exponentially with increasing $L$ when the frequency $\omega$ lies in the gap. Instead, we find that near the Dirac point $I\propto 1/L$. The $1/L$-scaling is reminiscent of diffusion through a disordered medium, but here it appears in the absence of any disorder inside the photonic crystal.

Such ``pseudo-diffusive'' scaling was discovered in Refs.\ \cite{Kat06,Two06} for electrical conduction through graphene (a two-dimensional carbon lattice with a Dirac point in the spectrum). Both the electronic and optical problems are governed by the same Dirac equation inside the medium, but the coupling to the outside space is different. In the electronic problem, the coupling can become nearly ideal for electrical contacts made out of heavily doped graphene \cite{Kat06,Two06}, or by suitably matching the Fermi energy in metallic contacts \cite{Sch06,Bla06}. An analogous freedom does not exist in the optical case.

The major part of our analysis is therefore devoted to the question how nonideal interfaces affect the dependence of $I$ on $\omega$ and $L$. Our conclusion is that
\begin{equation}
I/I_{0}=\Gamma_{0}W/L\label{IGamma}
\end{equation}
at the Dirac point, with $I_{0}$ the incident current per mode and $\Gamma_{0}$ an effective interface transparency. The properties of the interfaces determine the proportionality constant $\Gamma_{0}$, and they also determine whether $I$ as a function of $\omega$ has a minimum or a maximum near the Dirac point, but they leave the $1/L$-scaling unaffected.

In Sec.\ \ref{waveeq} we formulate the wave equations inside and outside the medium. The Helmholtz equation in free space is matched to the Dirac equation inside the photonic crystal by means of an interface matrix in Sec.\ \ref{matching}. This matrix could be calculated numerically, for a specific model for the termination of the crystal, but to arrive at general results we work with the general form of the interface matrix (constrained by the requirement of current conservation). The mode dependent transmission probability through the crystal is derived in Sec.\ \ref{transmission}. It depends on a pair of interface parameters for each of the two interfaces. In Sec.\ \ref{totaltrans} we then show that the extremal transmission near the Dirac point scales $\propto 1/L$ regardless of the values of these parameters. We conclude in Sec.\ \ref{conclude} with suggestions for experiments.

\section{Wave equations}
\label{waveeq}

\begin{figure}[tb]
\centerline{\includegraphics[width=0.9\linewidth]{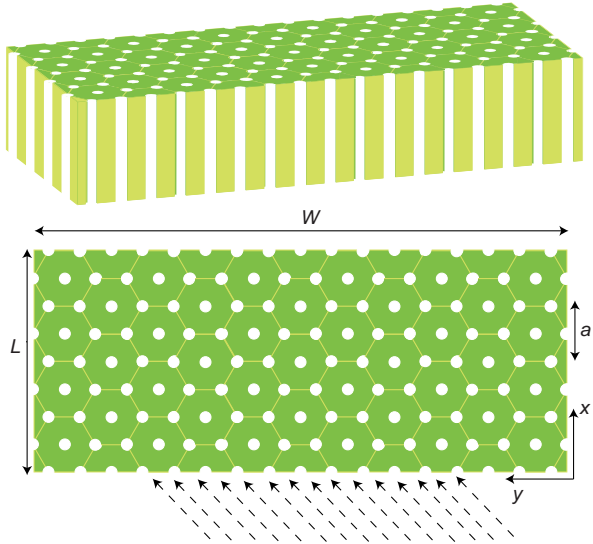}}
\caption{\label{fig_layout}
(Color online.) Photonic crystal formed by a dielectric medium perforated by parallel cylindrical holes on a triangular lattice (upper panel: front view; lower panel: top view). The dashed lines indicate the radiation incident on the armchair edge of the crystal, with the electric field polarized in the $z$-direction.
}
\end{figure}

We consider a two-dimensional photonic crystal consisting of a triangular or honeycomb lattice in the $x$-$y$ plane formed by cylindrical air-filled holes along the $z$-axis in a dielectric medium (see Fig.\ \ref{fig_layout}). The crystal has a width $W$ along the $y$-direction and a length $L$ along the $x$-direction, both dimensions being large compared to the lattice constant $a$. Monochromatic radiation (frequency $\omega$) is incident on the plane $x=0$, with the electric field $E(x,y)e^{i\omega t}$ polarized along the $z$-axis.

In the free space outside of the photonic crystal ($x<0$ and $x>L$) the Maxwell equations reduce to the Helmholtz equation
\begin{equation}
\left(\partial_{x}^{2}+\partial_{y}^{2}\right)E(x,y)+\frac{\omega^{2}}{c^{2}}E(x,y)=0.\label{Helmholtz}
\end{equation}
The mean (time averaged) photon number flux in the $x$-direction is given by \cite{Hou90}
\begin{equation}
j_{H}=\frac{\varepsilon_{0}c^{2}}{4i\hbar\omega^{2}}\left(E^{\ast}\frac{\partial E}{\partial x}-E\frac{\partial E^{\ast}}{\partial x}\right).\label{jHelmholtz}
\end{equation}

Inside the photonic crystal ($0<x<L$) the Maxwell equations reduce to the Dirac equation \cite{Rag06}
\begin{eqnarray}
\begin{pmatrix}
0&-iv_{D}(\partial_{x}-i\partial_{y})\\
-iv_{D}(\partial_{x}+i\partial_{y})&0
\end{pmatrix}
\begin{pmatrix}
\Psi_{1}\\
\Psi_{2}
\end{pmatrix}\nonumber\\
=(\omega-\omega_{D})
\begin{pmatrix}
\Psi_{1}\\
\Psi_{2}
\end{pmatrix},\label{Dirac}
\end{eqnarray}
for the amplitudes $\Psi_{1},\Psi_{2}$ of a doublet of two degenerate Bloch states at one of the corners of the hexagonal first Brillouin zone.

As explained by Raghu and Haldane \cite{Rag06,note1}, the modes at the six zone corners $\bm{K}_{p},\bm{K}'_{p}$ ($p=1,2,3$), which are degenerate for a homogeneous dielectric, are split by the periodic dielectric modulation into a pair of doublets at frequency $\omega_{D}$ and a pair of singlets at a different frequency. The first doublet and singlet have wave vectors at the first set of equivalent corners $\bm{K}_{p}$, while the second doublet and singlet are at $\bm{K}'_{p}$. Each doublet mixes and splits linearly forming a Dirac point as the wave vector is shifted by $\delta\bm{k}$ from a zone corner. The Dirac equation (\ref{Dirac}) gives the envelope field $\propto e^{i\delta\bm{k}\cdot\bm{r}}$ of one of these doublets.

The frequency $\omega_{D}$ and velocity $v_{D}$ in the Dirac equation depend on the strength of the periodic dielectric modulation, tending to $\omega_{D}=c'|\bm{K}_{p}|=c'|\bm{K}'_{p}|=4\pi c'/3a$ and $v_{D}=c'/2$ in the limit of weak modulation. (The speed of light $c'$ in the homogeneous dielectric is smaller than the free space value $c$.)

Eq.\ (\ref{Dirac}) may be written more compactly as
\begin{equation}
-iv_{D}(\nabla\cdot\bm{\sigma})\Psi=\delta\omega\Psi,\;\;\delta\omega\equiv \omega-\omega_{D},\label{Dirac2}
\end{equation}
in terms of the spinor $\Psi=(\Psi_{1},\Psi_{2})$ and the vector of Pauli matrices $\bm{\sigma}=(\sigma_{x},\sigma_{y})$. In the same notation, the
velocity operator for the Dirac equation is $v_{D}\bm{\sigma}$. The mean photon number flux $j_{D}$ in the $x$-direction is therefore given by
\begin{equation}
j_{D}=v_{D}\Psi^{\ast}\sigma_{x}\Psi=v_{D}(\Psi_{1}^{\ast}\Psi_{2}+\Psi_{2}^{\ast}\Psi_{1}).\label{jDirac}
\end{equation}

The termination of the photonic crystal in the $y$-direction introduces boundary conditions at the edges $y=0$ and $y=W$ which depend on the details of the edges, for example on edges being of zigzag,
armchair, or other type. For a wide and short crystal, $W\gg L$, these details become irrelevant and we may use periodic boundary conditions [$\Psi(x,0)=\Psi(x,W)$] for simplicity.

\section{Wave matching}
\label{matching}

\begin{figure}[tb]
\centerline{\includegraphics[width=0.9\linewidth]{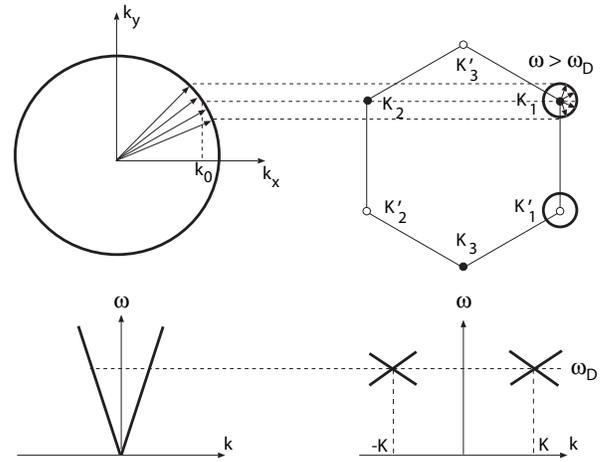}}
\caption{\label{fig_Brillouin}
Right panels: Hexagonal first Brillouin zone of the photonic crystal (top) and dispersion relation of the doublet near one of the zone corners (bottom). Filled and open dots distinguish the two sets of equivalent zone corners, centered at $\bm{K}_{p}$ and $\bm{K}'_{p}$, respectively. The small circles centered at the zone corners are the equal-frequency contours at a frequency $\omega$ just above the frequency $\omega_{D}$ of the Dirac point. Left panels: Equal-frequency contour in free space (top) and corresponding dispersion relation (bottom). A plane wave in free space with $k_{x}$ close to $k_{0}$ (arrows in the upper left panel) excites Bloch waves in the photonic crystal with $\bm{k}$ close to $\bm{K}_{1}$ and $\bm{K}_{2}$ (arrows in the upper right panel), as dictated by conservation of $k_{y}$ and $\omega$ (dotted horizontal lines).
}
\end{figure}

The excitation of modes near a Dirac point has been discussed by Notomi \cite{Not00}, in terms of a figure similar to Fig.\ \ref{fig_Brillouin}. Because the $y$-component of the wave vector is conserved across the boundary at $x=0$, the doublet near $\bm{K}_{1}=(K_{x},K_{y})$ or $\bm{K}_{2}=(-K_{x},K_{y})$ can only  be excited if the incident radiation has a wave vector $\bm{k}=(k_{x},k_{y})$ with $k_{y}$ near $K_{y}$. The conservation of $k_{y}$ holds up to translation by a reciprocal lattice vector. We will consider here the case of
$|\bm{k}|<|\bm{K}_{p}|$, where no coupling to $\bm{K}_{3}$ is
allowed. The actual radius of the equal frequency
contour in the
free space at $\omega = \omega_D$ will depend on a particular
photonic crystal realization.

The incident plane waves $E_{\rm incident}=E_{0}e^{i\bm{k}\cdot\bm{r}}$ in free space that excite Bloch waves at a frequency $\delta\omega=\omega-\omega_{D}$ have $k_{y}=K_{y}[1+{\cal O}(\delta\omega/\omega_{D})]$ and $k_{x}=k_{0}[1+{\cal O}(\delta\omega/\omega_{D})]$ with
\begin{equation}
k_{0}=\sqrt{(\omega_{D}/c)^{2}-K_{y}^{2}}.\label{k0def}
\end{equation}
For $\delta\omega\ll\omega_{D}$ we may therefore write the incident wave in the form
\begin{equation}
E_{\rm incident}(x,y)=E_{+}(x,y)e^{ik_{0}x+iK_{y}y},\label{Eindef}
\end{equation}
with $E_{+}$ a slowly varying function. Similarly, the reflected wave will have $k_{y}\approx K_{y}$ and $k_{x}\approx -k_{0}$, so that we may write it as
\begin{equation}
E_{\rm reflected}(x,y)=E_{-}(x,y)e^{-ik_{0}x+iK_{y}y},\label{Erdef}
\end{equation}
with $E_{-}$ slowly varying.

The orientation of the Brillouin zone shown in Fig.\ \ref{fig_Brillouin} corresponds to an armchair edge of the triangular lattice at $x=0$. For this orientation only one of the two inequivalent doublets is excited for a given $k_{y}$. (The other doublet at $\bm{K}'_{1}$,\ $\bm{K}'_{2}$ is excited for $-k_{y}$.) A $90^{\circ}$ rotation of the Brillouin zone would correspond to a zigzag edge. Then a linear combination of the two inequivalent doublets is excited near $k_{y}=0$. For simplicity, we will restrict ourselves here to the case shown in the figure of separately excitable doublets.

While the conservation of the wave vector component parallel to the boundary determines which modes in the photonic crystal are excited, it does not determine with what strength. For that purpose we need to match the solutions of the Helmholtz and Dirac equations at $x=0$. The matching should preserve the flux through the boundary, so it is convenient to write the flux in the same form at both sides of the boundary.

The photon number flux (\ref{jHelmholtz}) for the Helmholtz equation may be written in the same form as the flux (\ref{jDirac}) for the Dirac equation, by
\begin{subequations}
\label{jHelmholtz2}
\begin{eqnarray}
&&j_{H}=v_{H}{\cal E}^{\ast}\sigma_{x}{\cal E},\label{jHelmholtz2a}\\
&&v_{H}=\frac{\varepsilon_{0}c^{2}k_{0}}{4\hbar\omega^{2}},\;\;
{\cal E}=\begin{pmatrix}
E_{+}+E_{-}\\E_{+}-E_{-}
\end{pmatrix}.
\label{jHelmholtz2b}
\end{eqnarray}
\end{subequations}
(In the prefactor $k_{0}$ we have neglected corrections of order $\delta\omega/\omega_{D}$.) Flux conservation then requires
\begin{equation}
v_{H}{\cal E}^{\ast}\sigma_{x}{\cal  E}=v_{D}\Psi^{\ast}\sigma_{x}\Psi,\;\; {\rm at}\;\; x=0.\label{fluxconserve}
\end{equation}

The matching condition has the general form \cite{And89}
\begin{equation}
\Psi=(v_{H}/v_{D})^{1/2}M{\cal E},\;\;{\rm at}\;\; x=0.\label{Mdef}
\end{equation}
The flux conservation condition (\ref{fluxconserve}) implies that the transfer matrix $M$ should satisfy a generalized unitarity condition,
\begin{equation}
M^{-1}=\sigma_{x}M^{\dagger}\sigma_{x}.\label{symplectic}
\end{equation}
Eq.\ (\ref{symplectic}) restricts $M$ to a three-parameter form
\begin{equation}
M=e^{\gamma\sigma_{z}}e^{\beta\sigma_{y}}e^{i\alpha\sigma_{x}}\label{Malpha}
\end{equation}
(ignoring an irrelevant scalar phase factor). The real parameters $\alpha,\beta,\gamma$ depend on details of the boundary at the scale of the lattice constant --- they can not be determined from the Helmholtz or Dirac equations (the latter only holds on length scales $\gg a$).

We now show that the value of $\alpha$ becomes irrelevant close to the Dirac point. At the boundary the incident and reflected waves have the form
\begin{equation}
{\cal E}_{\rm incident}=E_{0}
\begin{pmatrix}1\\ 1\end{pmatrix},\;\;
{\cal E}_{\rm reflected}=rE_{0}
\begin{pmatrix}1\\ -1\end{pmatrix},\label{Einr}
\end{equation}
with $r$ the reflection coefficient, and $E_{0}\equiv E_{+}(0,y)$ a slowly varying function. Both ``spinors'' are eigenvectors of $\sigma_{x}$, hence the action of $e^{i\alpha\sigma_{x}}$ on ${\cal E}$ is simply a phase factor:
\begin{eqnarray}
&&M{\cal E}_{\rm incident}=e^{\gamma\sigma_{z}}e^{\beta\sigma_{y}}e^{i\alpha}{\cal E}_{\rm incident},\nonumber\\
&&M{\cal E}_{\rm reflected}=e^{\gamma\sigma_{z}}e^{\beta\sigma_{y}}e^{-i\alpha}{\cal E}_{\rm reflected}.\label{ME}
\end{eqnarray}
There is no need to determine the phase factor $e^{\pm i\alpha}$, since it has no effect on the reflection probability $|r|^{2}$.

A similar reasoning applies at the boundary $x=L$, where the matching condition reads
\begin{equation}
\Psi=(v_{H}/v_{D})^{1/2}M'{\cal E},\;\;{\rm at}\;\; x=L.\label{M2def}
\end{equation}
Flux conservation requires that $M'=e^{\gamma'\sigma_{z}}e^{\beta'\sigma_{y}}e^{i\alpha'\sigma_{x}}$, with real parameters $\alpha',\beta',\gamma'$. The value of $\alpha'$ is again irrelevant close to the Dirac point, because the spinor of the transmitted wave
\begin{equation}
{\cal E}_{\rm transmitted}=tE_{0}\begin{pmatrix}1\\ 1\end{pmatrix}\label{Et}
\end{equation}
(with $t$ the transmission coefficient) is an eigenvector of $\sigma_{x}$. So
\begin{equation}
M'{\cal E}_{\rm transmitted}=e^{\gamma'\sigma_{z}}e^{\beta'\sigma_{y}}e^{i\alpha'}{\cal E}_{\rm transmitted},\label{ME2}
\end{equation}
with a phase factor $e^{i\alpha'}$ that has no effect on the transmission probability $|t|^{2}$.

\section{Transmission probability}
\label{transmission}

We consider the case $W\gg L$ of a wide and short crystal, when we may use periodic boundary conditions at $y=0,W$ for the Bloch waves $\Psi\propto e^{i\delta\bm{k}\cdot\bm{r}}$. The transverse wave vector $\delta k_{y}$  is then discretized at $\delta k_{y}=2\pi n/W\equiv q_{n}$, with mode index $n=0,\pm 1,\pm 2,\pm 3,\ldots$. We seek the transmission amplitude $t_{n}$ of the $n$-th mode.

We first determine the transfer matrix $M_{n}(x,0)$ of the $n$-th mode $\Phi_{n}(x)e^{iq_{n}y}$ through the photonic crystal, defined by
\begin{equation}
\Phi_{n}(x)=M_{n}(x,0)\Phi_{n}(0).\label{Mndef}
\end{equation}
From the Dirac equation (\ref{Dirac2}) we obtain the differential equation
\begin{equation}
\frac{d}{dx}M_{n}(x,0)=
\left(\frac{i\delta\omega}{v_{D}}\sigma_{x}+q_{n}\sigma_{z}\right)M_{n}(x,0),\label{Mequation}
\end{equation}
with solution
\begin{equation}
M_{n}(x,0)=\cos k_{n}x+\frac{\sin k_{n}x}{k_{n}}\left(\frac{i\delta\omega}{v_{D}}\sigma_{x}+q_{n}\sigma_{z}\right).\label{Mresult}
\end{equation}
We have defined the longitudinal wave vector
\begin{equation}
k_{n}=\sqrt{(\delta\omega/v_{D})^{2}-q_{n}^{2}}.\label{kndef}
\end{equation}

The total transfer matrix through the photonic crystal, including the contributions (\ref{Mdef})  and (\ref{M2def}) from the interfaces at $x=0$ and $x=L$, is
\begin{equation}
{\cal M}=M'^{-1}M_{n}(L,0)M.\label{calMdef}
\end{equation}
It determines the transmission amplitude by
\begin{eqnarray}
{\cal M}\begin{pmatrix}
1+r_{n}\\ 1-r_{n}
\end{pmatrix}=
\begin{pmatrix}t_{n}\\ t_{n}\end{pmatrix}&\Rightarrow&
\begin{pmatrix}
1-r_{n}\\ 1+r_{n}
\end{pmatrix}={\cal M}^{\dagger}\begin{pmatrix}t_{n}\\ t_{n}\end{pmatrix}\nonumber\\
&\Rightarrow&\frac{1}{t_{n}}=\frac{1}{2}
\sum_{i=1}^{2}\sum_{j=1}^{2}
{\cal M}^{\ast}_{ij},\label{tnMrelation}
\end{eqnarray}
where we have used the current conservation relation ${\cal M}^{-1}=\sigma_{x}{\cal M}^{\dagger}\sigma_{x}$.

The general expression for the transmission probability $T_{n}=|t_{n}|^{2}$ is rather lengthy, but it simplifies in the case that the two interfaces at $x=0$ and $x=L$ are related by a reflection symmetry. For a photonic crystal that has an axis of symmetry at $x=L/2$ both $\Phi(x)$ and $\sigma_{y}\Phi(L-x)$ are solutions at the same frequency. This implies for the transfer matrix the symmetry relation
\begin{eqnarray}
\sigma_{y}{\cal M}\sigma_{y}&=&{\cal M}^{-1}\Rightarrow
\sigma_{y}M'\sigma_{y}=M\nonumber\\
&\Rightarrow&\beta'=\beta,\;\;\gamma'=-\gamma,\label{MprimeMrelation}
\end{eqnarray}
and we obtain
\begin{widetext}
\begin{eqnarray}
\frac{1}{T_{n}}&=&\left(\frac{\delta\omega\sin k_{n}L}{v_{D}k_{n}}\cosh 2\beta-\cos k_{n}L\sinh 2\beta\sinh 2\gamma-\frac{q_{n}\sin k_{n}L}{k_{n}}\sinh 2\beta\cosh 2\gamma\right)^{2}\nonumber\\
&&\mbox{}+
\left(\cos k_{n}L\cosh 2\gamma+\frac{q_{n}\sin k_{n}L}{k_{n}}\sinh 2\gamma\right)^{2}.\label{Tresult}
\end{eqnarray}
\end{widetext}
For an ideal interface (when $\beta=0=\gamma$) we recover the transmission probability of Ref.\ \cite{Two06}.

At the Dirac point, where $\delta\omega=0\Rightarrow k_{n}=iq_{n}$, Eq.\ (\ref{Tresult}) reduces further to
\begin{equation}
\frac{1}{T_{n}}=\cosh^{2}(q_{n}L+2\gamma)+\sinh^{2}2\beta\sinh^{2}(q_{n}L+2\gamma).\label{TresultDirac}
\end{equation}
More generally, for two arbitrary interfaces, the transmission probability at the Dirac point takes the form
\begin{eqnarray}
&&\frac{1}{T_{n}}=\cosh^{2}(\beta-\beta')\cosh^{2}\xi_{n}+\sinh^{2}(\beta+\beta')\sinh^{2}\xi_{n},\nonumber\\
&&\xi_{n}=q_{n}L+\gamma-\gamma'.\label{TresultDiracgeneral}
\end{eqnarray}
While the individual $T_{n}$'s depend on $\gamma$ and $\gamma'$, this dependence drops out in the total transmission $\sum_{n}T_{n}$.

\section{Photon current}
\label{totaltrans}

The transmission probabilities determine the time averaged photon current $I$ at frequency $\omega_{D}+\delta\omega$ through the photonic crystal,
\begin{equation}
I(\delta\omega)=I_{0}\sum_{n=-\infty}^{\infty}T_{n}(\delta\omega),\label{barI}
\end{equation}
where $I_{0}$ is the incident photon current per mode. The sum over $n$ is effectively cut off at $|n|\sim W/L\gg 1$, because of the exponential decay of the $T_{n}$'s for larger $|n|$. This large number of transverse modes excited in the photonic crystal close to the Dirac point corresponds in free space to a narrow range $\delta\phi\simeq a/L\ll 1$ of angles of incidence. We may therefore assume that the incident radiation is isotropic over this range of angles $\delta\phi$, so that the incident current per mode $I_{0}$ does not depend on $n$.

\begin{figure}[tb]
\centerline{\includegraphics[width=0.9\linewidth]{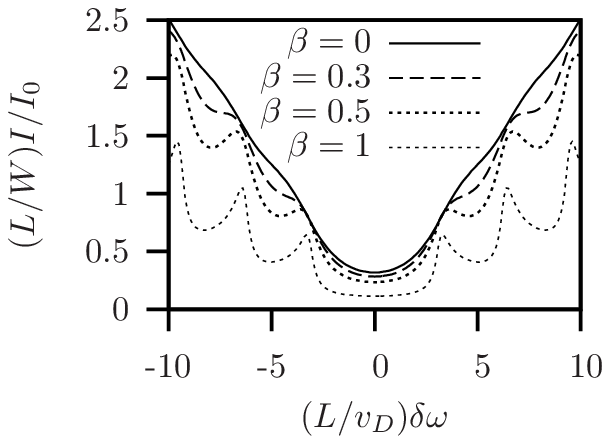}}\medskip

\centerline{\includegraphics[width=0.9\linewidth]{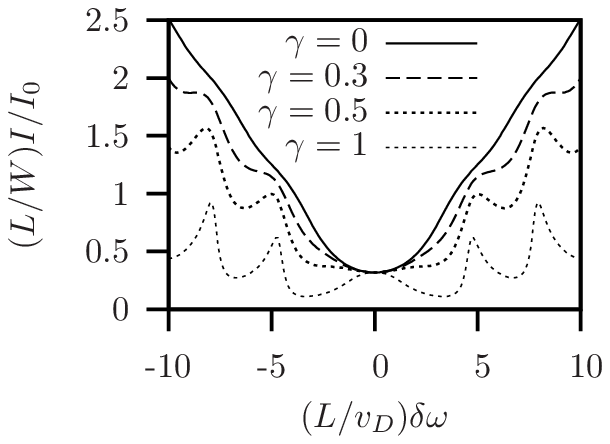}}
\caption{\label{fig_I}
Frequency dependence of the transmitted current, for interface parameters $\beta'=\beta$, $\gamma'=-\gamma$. In the top panel we take $\gamma=0$ and vary $\beta$, while in the bottom panel we take $\beta=0$ and vary $\gamma$. The solid curves ($\beta=\gamma=0$) correspond to maximal coupling of the photonic crystal to free space. The curves are calculated from Eqs.\ (\ref{Tresult}) and (\ref{barI}), in the regime $W/L\gg 1$ where the sum over modes may be replaced by an integration over transverse wave vectors.
}
\end{figure}

\begin{figure}[tb]
\centerline{\includegraphics[width=0.9\linewidth]{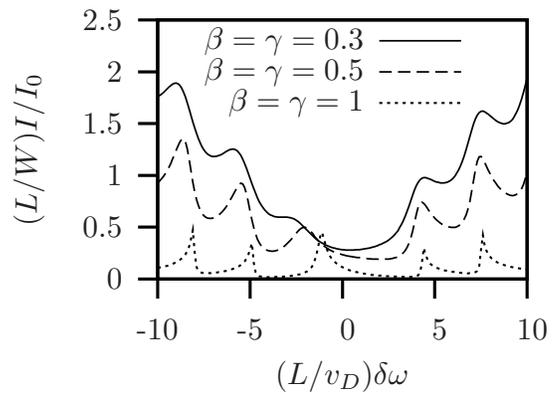}}
\caption{\label{fig_I2}
Same as Fig.\ \ref{fig_I}, for $\beta$ and $\gamma$ both nonzero.}
\end{figure}

Since $W/L\gg 1$ the sum over modes may be replaced by an integration over wave vectors, $\sum_{n=-\infty}^{\infty}\rightarrow(W/2\pi)\int_{-\infty}^{\infty}dq_{n}$. The resulting frequency dependence of the photon current around the Dirac frequency is plotted in Figs.\ \ref{fig_I} and \ref{fig_I2}, for several values of the interface parameters. As we will now discuss, the scaling with the separation $L$ of the interfaces is fundamentally different close to the Dirac point than it is away from the Dirac point.

Substitution of Eq.\ (\ref{TresultDiracgeneral}) into Eq.\ (\ref{barI}) gives the photon current at the Dirac point,
\begin{eqnarray}
&& I(\delta\omega=0)=I_{0}\Gamma_{0}\frac{W}{L},\nonumber\\
&&\Gamma_{0}=\frac{\arctan[\sinh(\beta+\beta')/\cosh(\beta-\beta')]}{\pi\sinh(\beta+\beta')\cosh(\beta-\beta')},\label{ItotalDirac}
\end{eqnarray}
independent of the parameters $\gamma,\gamma'$. For two ideal interfaces we reach the limit
\begin{equation}
\lim_{\beta,\beta'\rightarrow 0}I(\delta\omega=0)/I_{0}=\frac{1}{\pi}\frac{W}{L},\label{Itotalideal}
\end{equation}
in agreement with Refs.\ \cite{Kat06,Two06}. Eq.\ (\ref{ItotalDirac}) shows that, regardless of the transparency of the interfaces at $x=0$ and $x=L$, the photon current at the Dirac point is inversely proportional to the separation $L$ of the interfaces (as long as $a\ll L\ll W$).

As seen in Figs.\ \ref{fig_I} and \ref{fig_I2}, the photon current at the Dirac point has 
an extremum (minimum or maximum) when either $\gamma$ or $\beta$ are 
equal to zero. If the interface parameters $\beta,\gamma$ are both nonzero, then the extremum is displaced from the Dirac point by a frequency shift $\delta\omega_{c}$. The photon current $I(\delta\omega_{c})$ at the extremum remains inversely proportional to $L$ as in Eq.\ (\ref{ItotalDirac}), with a different proportionality constant $\Gamma_{0}$ (which now depends on both $\beta$ and $\gamma$).

The $1/L$-scaling of the photon current applies to a frequency interval $|\delta\omega|\lesssim v_{D}/L$ around the Dirac frequency $\omega_{D}$. For $|\delta\omega|\gg v_{D}/L$ the photon current approaches the $L$-independent value
\begin{equation}
I_{\infty}=I_{0}\Gamma\frac{W\delta\omega}{\pi v_{D}},\label{Itotallargeomega}
\end{equation}
with rapid oscillations around this limiting value. The effective interface transmittance $\Gamma$ is a rather complicated function of the interface parameters $\beta,\beta',\gamma,\gamma'$. It is still somewhat smaller than unity even for maximal coupling of the photonic crystal to free space ($\Gamma=\pi/4$ for $\beta=\gamma=0$).

\section{Conclusion}
\label{conclude}

While several experiments \cite{Cas97,Ye04} have studied two-dimensional photonic crystals with a honeycomb or triangular lattice, the emphasis has been on the frequency range where the band structure has a true gap, rather than on frequencies near the Dirac point. Recent experiments on electronic conduction near the Dirac point of graphene have shown that this singularity in the band structure offers a qualitatively new transport regime \cite{Gei07}. Here we have explored the simplest optical analogue, the pseudo-diffusive transmission extremum near the Dirac point of a photonic crystal. We believe that photonic crystals offer a particularly clean and controlled way to test this prediction experimentally. The experimental test in the electronic case is severely hindered by the difficulty to maintain a homogeneous electron density throughout the system \cite{Cas07}. No such difficulty exists in a photonic crystal.

If this experimental test is successful, there are other unusual effects at the Dirac point waiting to be observed. For example, disorder has been predicted to increase --- rather than decrease --- the transmission at the Dirac point \cite{Tit06,Ryc06,Ost07}. Photonic crystals could provide an ideal testing ground for these theories.

\acknowledgments

We have benefited from discussions with A. R. Akhmerov, Ya.\ M. Blanter, and M. de Dood. This research was supported by the Dutch Science Foundation NWO/FOM.

\end{document}